\documentclass[12pt]{article}
\renewcommand{\theequation}{\arabic{section}.\arabic{equation}}

\font\oneeight=cmr10 at 18pt

\newcommand{\vTm}{\vphantom{\mbox{\oneeight I}}}

\begin{document}



\def\a{\alpha}
\def\b{\beta}
\def\d{\delta}
\def\e{\epsilon}
\def\g{\gamma}
\def\h{\mathfrak{h}}
\def\k{\kappa}
\def\l{\lambda}
\def\o{\omega}
\def\p{\wp}
\def\r{\rho}
\def\t{\tau}
\def\s{\sigma}
\def\z{\zeta}
\def\x{\xi}
 \def\A{{\cal{A}}}
 \def\B{{\cal{B}}}
 \def\C{{\cal{C}}}
 \def\D{{\cal{D}}}
\def\G{\Gamma}
\def\K{{\cal{K}}}
\def\O{\Omega}
\def\L{\Lambda}
\def\f{E_{\tau,\eta}(sl_2)}
\def\E{E_{\tau,\eta}(sl_n)}
\def\Zb{\mathbf{Z}}
\def\Cb{\mathbf{C}}
\def\T{\mathcal{T}}

\def\R{\overline{R}}

\def\beq{\begin{equation}}
\def\eeq{\end{equation}}
\def\bea{\begin{eqnarray}}
\def\eea{\end{eqnarray}}
\def\ba{\begin{array}}
\def\ea{\end{array}}
\def\no{\nonumber}
\def\le{\langle}
\def\re{\rangle}
\def\lt{\left}
\def\rt{\right}

\newtheorem{Theorem}{Theorem}
\newtheorem{Definition}{Definition}
\newtheorem{Proposition}{Proposition}
\newtheorem{Lemma}{Lemma}
\newtheorem{Corollary}{Corollary}
\newcommand{\proof}[1]{{\bf Proof. }
        #1\begin{flushright}$\Box$\end{flushright}}

\baselineskip=20pt

\newfont{\elevenmib}{cmmib10 scaled\magstep1}
\newcommand{\preprint}{
   \begin{flushright}\normalsize  \sf
     {\tt hep-th/0411190} \\ November 2004
   \end{flushright}}
\newcommand{\Title}[1]{{\baselineskip=26pt
   \begin{center} \Large \bf #1 \\ \ \\ \end{center}}}
\newcommand{\Author}{\begin{center}
   \large \bf
Wen-Li Yang${}^{a,b}$
 ~ and~Yao-Zhong Zhang${}^b$\end{center}}
\newcommand{\Address}{\begin{center}

     ${}^a$ Institute of Modern Physics, Northwest University
     Xian 710069, P.R. China\\
     ${}^b$ Department of Mathematics, The University of Queensland,
     Brisbane 4072, Australia\\
     E-mail: wenli@maths.uq.edu.au, yzz@maths.uq.edu.au
   \end{center}}
\newcommand{\Accepted}[1]{\begin{center}
   {\large \sf #1}\\ \vspace{1mm}{\small \sf Accepted for Publication}
   \end{center}}

\preprint
\thispagestyle{empty}
\bigskip\bigskip\bigskip

\Title{Exact solution of the $A^{(1)}_{n-1}$ trigonometric vertex
model with non-diagonal open boundaries} \Author

\Address
\vspace{1cm}

\begin{abstract}
The $A^{(1)}_{n-1}$ trigonometric vertex model with {\it generic
non-diagonal\/} boundaries is studied. The double-row transfer
matrix of the model is diagonalized by algebraic Bethe ansatz
method in terms of the intertwiner and the corresponding
face-vertex relation. The eigenvalues and the corresponding Bethe
ansatz equations are obtained.

\vspace{1truecm} \noindent {\it PACS:} 03.65.Fd; 05.30.-d; 05.50+q

\end{abstract}
\newpage
\section{Introduction}
\label{intro} \setcounter{equation}{0}

Two-dimensional integrable models have traditionally been solved
by imposing periodic boundary conditions. For such bulk systems,
the quantum Yang-Baxter equation (QYBE)\bea
R_{12}(u_1-u_2)R_{13}(u_1-u_3)R_{23}(u_2-u_3)=
R_{23}(u_2-u_3)R_{13}(u_1-u_3)R_{12}(u_1-u_2), \label{QYB} \eea
leads to families of commuting row-to-row {\it transfer matrices}
which may be diagonalized  by the quantum inverse scattering
method (QISM) (or algebraic Bethe ansatz) \cite{Kor93}.

Not all boundary conditions are compatible with integrability in
the bulk. The bulk integrability is only preserved  when one
imposes certain boundary conditions. In \cite{Skl88}, Sklyanin
developed the boundary QISM, which may be used to described
integrable systems on a finite interval with independent boundary
conditions at each end. This boundary QISM uses the new algebraic
structure, the reflection equation (RE) algebra. The solutions to
the RE and its dual are called boundary K-matrices which in turn
give rise to boundary conditions compatible with the integrability
of the bulk model \cite{Skl88}-\cite{Gho94}.

The boundary QISM has been applied to diagonalize the double-row
transfer matrices of various integrable models with non-trivial
boundary conditions mostly corresponding to the diagonal
K-matrices. However, the problem of diagonalizing the double-row
transfer matrix for general {\it non-diagonal\/} K-matrices has
been  long-standing for trigonometric integrable models. To our
knowledge, the only exception is the spin-$\frac{1}{2}$ XXZ (or
$A^{(1)}_{1}$ ) model with non-diagonal K-matrices which was
solved  recently by fusion hierarchy of the transfer matrix with
the anisotropy value being the roots of unity \cite{Nep02,Nep03},
the algebraic Bethe ansatz  \cite{Cao03,Yan043} and the coordinate
Bethe ansatz  \cite{Gie04}. The fundamental difficulty is that the
usual highest-weight state  which is the pseudo-vacuum (or
reference state) for the models with periodic boundary condition
or  boundary conditions specified by diagonal K-matrices is {\it
no longer\/} the pseudo-vacuum  on which the Bethe ansatz analysis
is based.

In a very recent work \cite{Yan042}, we constructed a class of
non-diagonal solutions to  the RE for the trigonometric
$A^{(1)}_{n-1}$  vertex model by the intertwiner-matrix approach.
The {\it non-diagonal\/} K-matrices we found  can be expressed in
terms of the intertwiner-matrices and diagonal face-type
K-matrices. In the present paper, we solve the trigonometric
$A^{(1)}_{n-1}$ vertex models with boundary conditions given by
the non-diagonal K-matrices in \cite{Yan042}. We construct the
pseudo-vacuum and diagonalize the corresponding double-row
transfer matrix by the generalized QISM developed in \cite{Yan04}.

This paper is organized as follows. In section 2, we introduce our
notation and some  basic ingredients. In section 3, we introduce
the intertwiner-matrix which satisfies the face-vertex
correspondence relation between the two R-matrices $R(u)$ and
$W(u)$. Through the {\it magic\/} intertwiner vectors, in section
4, we transform the model from the original vertex picture into
its ``face" picture. After succeeding in constructing the
pseudo-vacuum state, we apply the algebraic Bethe ansatz method to
diagonalize the transfer matrices of the boundary model. Section 5
is for conclusions. Some detailed technical calculations are given
in Appendices A-C.

\section{$A^{(1)}_{n-1}$ trigonometric vertex model and integrable boundary conditions}
 \label{RE} \setcounter{equation}{0}

Let us fix a positive integer $n$ ($n\geq 2$) and a generic
complex number $\eta$, and  $R(u)\in End(\Cb^n\otimes\Cb^n)$ be
the trigonometric solution to the $A^{(1)}_{n-1}$ type QYBE given
by \cite{Che80,Per81,Baz91} \bea
\hspace{-0.5cm}R(u)=\sum_{\a=1}^{n}R^{\a\a}_{\a\a}(u)E_{\a\a}\otimes
E_{\a\a} +\sum_{\a\ne \b}\lt\{R^{\a\b}_{\a\b}(u)E_{\a\a}\otimes
E_{\b\b}+ R^{\b\a}_{\a\b}(u)E_{\b\a}\otimes
E_{\a\b}\rt\},\label{R-matrix} \eea where $E_{ij}$ is the matrix
with elements $(E_{ij})^l_k=\d_{jk}\d_{il}$. The coefficient
functions are \bea
R^{\a\b}_{\a\b}(u)&=&\lt\{\begin{array}{cc}\frac{\sin(u)}{\sin(u+\eta)}\,e^{-i\eta},&
\a>\b,\\[6pt]1,&\a=\b,\\[6pt]\frac{\sin(u)}{\sin(u+\eta)}\,e^{i\eta},&
\a<\b,\end{array}\rt.,\label{Elements1}\\[6pt]
R^{\b\a}_{\a\b}(u)&=&\lt\{\begin{array}{cc}\frac{\sin(\eta)}{\sin(u+\eta)}\,e^{iu},&
\a>\b,\\[6pt]1,&\a=\b,\\[6pt]\frac{\sin(\eta)}{\sin(u+\eta)}\,e^{-iu},&
\a<\b,\end{array}\rt..\label{Elements2}\eea One can check that the
R-matrix satisfies the following unitarity, crossing-unitarity and
quasi-classical relations:\begin{eqnarray}
 &&\hspace{-1.5cm}\mbox{
 Unitarity}:\hspace{42.5mm}R_{12}(u)R_{21}(-u)= {\rm id},\label{Unitarity}\\
 &&\hspace{-1.5cm}\mbox{
 Crossing-unitarity}:\quad
 R^{t_2}_{12}(u)M_2^{-1}R_{21}^{t_2}(-u-n\eta)M_2
 = \frac{\sin(u)\sin(u+n\eta)}{\sin(u+\eta)\sin(u+n\eta-\eta)}\,\mbox{id},
 \label{crosing-unitarity}\\
 &&\hspace{-1.5cm}\mbox{ Quasi-classical
 property}:\hspace{22.5mm}\, R_{12}(u)|_{\eta\rightarrow 0}= {\rm
id}.\label{quasi}
\end{eqnarray}
Here $R_{21}(u)=P_{12}R_{12}(u)P_{12}$ with $P_{12}$ being the
usual permutation operator and $t_i$ denotes the transposition in
the $i$-th space, and $\eta$ is the so-called crossing paramter.
The crossing matrix $M$ is a diagonal $n\times n$ matrix with
elements \bea
M_{\a\b}=M_{\a}\d_{\a\b},~~M_{\a}=e^{-2i\a\eta},~\a=1,\ldots,n.\label{C-Matrix}\eea
Here and below we adopt the standard notation: for any matrix
$A\in {\rm End}(\Cb^n)$, $A_j$ is an embedding operator in the
tensor space $\Cb^n\otimes \Cb^n\otimes\cdots$, which acts as $A$
on the $j$-th space and as an identity on the other factor spaces;
$R_{ij}(u)$ is an embedding operator of R-matrix in the tensor
space, which acts as an identity on the factor spaces except for
the $i$-th and $j$-th ones.

One introduces  the ``row-to-row" monodromy matrix $T(u)$, which
is an $n\times n$ matrix with elements being operators acting  on
$(\Cb^n)^{\otimes N}$  \begin{eqnarray}
T(u)=R_{01}(u+z_1)R_{02}(u+z_2)\cdots
R_{0N}(u+z_N).\label{T-matrix}\end{eqnarray} Here
$\{z_i|i=1,\ldots, N\}$ are arbitrary free complex parameters
which are usually called inhomogeneous parameters. With the help
of the QYBE (\ref{QYB}), one can show that $T(u)$ satisfies the
so-called ``RLL" relation
\begin{eqnarray}
R_{12}(u-v)T_1(u)T_2(v)=T_2(v)T_1(u)R_{12}(u-v).\label{Relation1}\end{eqnarray}

Integrable open chains can be constructed as follows \cite{Skl88}.
Let us introduce the K-matrix $K^-(u)$ which gives rise to an
integrable boundary condition on the right boundary. $K^-(u)$
satisfies  the RE
 \begin{eqnarray}
 &&R_{12}(u_1-u_2)K^-_1(u_1)R_{21}(u_1+u_2)K^-_2(u_2)\no\\
  &&~~~~~~=
 K^-_2(u_2)R_{12}(u_1+u_2)K^-_1(u_1)R_{21}(u_1-u_2).\label{RE-V}
\end{eqnarray}
For models with open boundaries, instead of the standard
``row-to-row" monodromy matrix $T(u)$ (\ref{T-matrix}), one needs
the  ``double-row" monodromy matrix $\mathbf{T}(u)$
\begin{eqnarray}
 \mathbf{T}(u)=T(u)K^-(u)T^{-1}(-u).\label{Mon-V-1}
\end{eqnarray}
Using (\ref{Relation1}) and (\ref{RE-V}), one can prove that
$\mathbf{T}(u)$ satisfies
\begin{eqnarray}
 R_{12}(u_1-u_2)\mathbf{T}_1(u_1)R_{21}(u_1+u_2)
  \mathbf{T}_2(u_2)=
 \mathbf{T}_2(u_2)R_{12}(u_1+u_2)\mathbf{T}_1(u_1)R_{21}(u_1-u_2).
 \label{Relation-Re}
\end{eqnarray}
In order to construct the {\it double-row transfer matrices},
besides the RE, one  needs another K-matrix $K^+(u)$ which gives
integrable boundary condition on the left boundary. The K-matrix
$K^+(u)$ satisfies the dual RE \cite{Skl88,Mez91,Yan042} \bea
&&R_{12}(u_2-u_1)K^+_1(u_1)\,M_1^{-1}\,R_{21}(-u_1-u_2-n\eta)\,M_1\,K^+_2(u_2)\no\\
&&\qquad\quad=
M_1\,K^+_2(u_2)R_{12}(-u_1-u_2-n\eta)\,M_1^{-1}\,K^{+}_1(u_1)R_{21}(u_2-u_1).
\label{DRE-V1}\eea Different  integrable boundary conditions are
described by different solutions $K^{-}(u)$ ($K^{+}(u)$) to the
(dual) RE \cite{Skl88, Gho94}. In this paper, we consider the
non-diagonal solutions $K^{\pm}(u)$ obtained in  \cite{Yan042},
which are respectively given by
 \bea K^-(u)^s_t&=&\sum_{i=1}^n
k_i(u)\phi^{(s)}_{\l,\l-\e_{i}}(u)
\bar{\phi}^{(t)}_{\l,\l-\e_{i}}(-u), \label{K-matrix}\\
K^+(u)^s_t&=&\sum_{i=1}^n
\tilde{k}_i(u)\phi^{(s)}_{\l',\l'-\e_{i}}(-u)
\tilde{\phi}^{(t)}_{\l',\l'-\e_{i}}(u). \label{DK-matrix}\eea Here
$\{k_i(u)|i=1,\ldots,n\}$ and $\{\tilde{k}_i(u)|i=1,\ldots,n\}$
are
\bea k_j(u)&=&\lt\{\begin{array}{ll}1,&1\leq j\leq l,\\[4pt]
\frac{\sin(\xi-u)}{\sin(\xi+u)}e^{-2iu},&l+1\leq j\leq
n,\end{array}\rt.\label{K-matrix1}\\
\tilde{k}_j(u)&=&\lt\{\begin{array}{ll}e^{-2i(j\eta)},&1\leq j\leq l',\\[4pt]
\frac{\sin(\bar{\xi}+u+\frac{n}{2}\eta)}{\sin(\bar{\xi}-u-\frac{n}{2}\eta)}
e^{2i(u+\frac{n-2j}{2}\eta)},&l'+1\leq j\leq
n,\end{array}\rt.\label{DK-matrix1}\eea where $l$ and $l'$ are
positive integers such that $1\leq l\leq n$, $1\leq l'\leq n$. In
(\ref{K-matrix}) and (\ref{DK-matrix}), $\phi$, $\bar{\phi}$ and
$\tilde{\phi}$ are intertwiners which will be specified in section
3. The K-matrix $K^-(u)$ (resp. $K^+(u)$) depends on a {\it
discrete\/} parameter $l$ (resp. $l'$) and {\it continuous\/}
parameters $\xi$, $\{\l_i\}$ (resp. $\bar{\xi}$, $\{\l'_i\}$) and
$\rho$ (whose dependence is through the definition of the
intertwiner-matrix (\ref{In-matrix}) below). Here and throughout
we use the convention: associated with the boundary parameters
$\{\l_i\}$ (resp. $\{\l'_i\}$) let us introduce a vector
$\l=\sum_{i=1}^n\l_i\e_i $ (resp. $\l'=\sum_{i=1}^n\l'_i\e_i$),
where $\{\e_i|\,i=1,\ldots,n\}$ is the orthonormal basis of the
vector space $\Cb^n$ such that $\langle \e_i,\e_j\rangle=\d_{ij}$.

Some remarks are in order. Our bulk R-matrix is different from
those used in  \cite{Lim02,Aba95,Yam02} (see (\ref{Elements1})).
For $n=3$ case, after a similarity transformation by a
spectral-independent {\it diagonal\/} matrix, our K-matrix
$K^-(u)$ has the same number of the boundary parameters as that of
\cite{Yam02} and one more boundary parameter than that of
\cite{Lim02,Aba95}. For generic $n$ $(n>3)$, our K-matrix $K^-(u)$
has many more boundary parameters than that given in \cite{Lim02}.

Let us emphasize  that a further restriction \bea
\l'+\sum_{k=1}^N\e_{j_k}=\l,\label{Restriction}\eea where
$\{j_k|k=1,\ldots,N\}$ are positive integers such that $ 2\leq
j_k\leq n$, is necessary for the the application of the algebraic
Bethe ansatz method in section 4. Hereafter, we shall consider
only the above case.

The {\it double-row transfer matrix\/} of  the inhomogeneous model
associated with the R-matrix (\ref{R-matrix})-(\ref{Elements2})
with open boundary specified by the K-matrices $K^{\pm}(u)$
(\ref{K-matrix})-(\ref{Restriction}) is given by
\begin{eqnarray}
\tau(u)=tr(K^+(u)\mathbf{T}(u)).\label{trans}
\end{eqnarray} The commutativity of the transfer matrices
\begin{eqnarray}
 [\tau(u),\tau(v)]=0,\label{Com-2}
\end{eqnarray}
follows as a consequence of (\ref{QYB}),
(\ref{Unitarity})-(\ref{crosing-unitarity}) and
(\ref{Relation-Re})-(\ref{DRE-V1}). This ensures the integrability
of the inhomogeneous  model with open boundary. The aim of this
paper is to find the common eigenvectors and the corresponding
eigenvalues of the transfer matrix (\ref{trans}) with generic
non-diagonal K-matrices  $K^{\pm}(u)$ given by
(\ref{K-matrix})-(\ref{Restriction}).

\section{Intertwining vectors and the associated face-vertex correspondence relations}
 \label{IRF} \setcounter{equation}{0}

For a vector $m\in \Cb^n$, set \bea m_i=\langle m,\e_i\rangle,
~~|m|=\sum_{k=1}^nm_k,~~i=1,\ldots,n. \label{Def1}\eea Let us
introduce $n$ intertwining vectors (intertwiners)
$\{\phi_{m,m-\e_j}(u)|\, j=1,\ldots,n\}$. Each
$\phi_{m,m-\e_j}(u)$ is an $n$-component column vector whose
$\a$-th elements are $\{\phi^{(\a)}_{m,m-\e_j}(u)\}$. The $n$
intertwiners form an $n\times n$ matrix (in which $j$ and $\a$
stand for the column and the row indices respectively), called the
intertwiner-matrix, with the non-vanishing matrix elements being
\bea \lt(\begin{array}{ccccccc}e^{i\eta f_1(m)} &&&&&&e^{i\eta
F_n(m)+\rho}e^{2iu}\\e^{i\eta F_1(m)}&e^{i\eta f_2(m)}&&&&&\\
&e^{i\eta F_2(m)}&\ddots&&&&\\&&\ddots&e^{i\eta
f_j(m)}&&&\\&&&e^{i\eta F_j(m)}&\ddots&&\\&&&&\ddots&e^{i\eta
f_{n-1}(m)}&\\&&&&&e^{i\eta F_{n-1}(m)}&e^{i\eta f_n(m)}
\end{array}\rt).\label{In-matrix}\eea Here $\rho$  is a complex constant
with regard to $u$ and $m$,  and $\{f_i(m)|i=1,\ldots,n\}$ and
$\{F_{i}(m)|i=1,\ldots,n\}$ are linear functions of $m$:\bea
f_i(m)&=&\sum_{k=1}^{i-1}m_k-m_i-\frac{1}{2}|m|,~~i=1,\ldots,n,\label{function1}\\
F_{i}(m)&=&\sum_{k=1}^{i}m_k-\frac{1}{2}|m|,~~i=1,\ldots,n-1,\label{function2}\\
F_n(m)&=&-\frac{3}{2}|m|.\label{function3}\eea From the above
intertwiner-matrix, one may derive  the following {\it face-vertex
correspondence relation\/} \cite{Yan042}:\bea &&R_{12}(u_1-u_2)
\phi_{m,m-\e_i}(u_1)\otimes
\phi_{m-\e_i,m-\e_i-\e_j}(u_2) \no\\
&&~~~~=\sum_{k,l}W^{kl}_{ij}(u_1-u_2)
\phi_{m-\e_{l},m-\e_{l}-\e_{k}}(u_1)\otimes
\phi_{m,m-\e_{l}}(u_2).\label{Face-vertex}\eea Here the
non-vanishing elements of $\{W(u)^{kl}_{ij}\}$ are \bea
W^{jj}_{jj}(u)&=&1,~~W^{jk}_{jk}(u)=\frac{\sin(u)}{\sin(u+\eta)},
~~{\rm for}\,\,j\neq k,\label{W-elements-1}\\[6pt]
W^{kj}_{jk}(u)&=&\lt\{\begin{array}{cc}\frac{\sin(\eta)}
{\sin(u+\eta)}\,e^{iu},&
j>k,\\[6pt]\frac{\sin(\eta)}{\sin(u+\eta)}\,e^{-iu},&
j<k,\end{array}\rt.~~{\rm for}~j\neq k.\label{W-elements-2}\eea
Associated with $\{W(u)^{kl}_{ij}\}$, one may introduce ``{\it
face"\/} type R-matrix $W(u)$ \bea W(u)=\sum_{i,j,k,l}
W^{kl}_{ij}(u)E_{ki}\otimes E_{lj}.\label{W-matrix}\eea Some
remarks are in order. The ``face" type R-matrix $W(u)$ does not
depend on the face type parameter $m$, in contrast to
 the $\Zb_n$ elliptic case \cite{Bel81,Jim87}.
 It follows that $W(u)$ and $R(u)$
 satisfy the same QYBE, i.e. $W(u)$ obeys the usual (vertex type) QYBE rather than
 the {\it dynamical\/} one \cite{Hou03,Yan04}.

Noting that \bea
\sum_{i=1}^nf_i(m)=\sum_{i=1}^nF_i(m)=\sum_{k=1}^{n}\frac{n-2(k+1)}{2}m_k,\eea
one can show that the determinant of the intertwiner matrix
(\ref{In-matrix}) is \bea {\rm
Det}\lt(\phi^{(\a)}_{m,m-\e_j}(u)\rt)=e^{i\eta\sum_{k=1}^{n}\frac{n-2(k+1)}{2}m_k}\,
(1-(-1)^ne^{2iu+\rho}).\label{Det}\eea For a generic $\rho\in\Cb$
this determinant is non-vanishing and thus the inverse of
(\ref{In-matrix}) exists. This fact allows us to introduce other
types of intertwiners $\bar{\phi}$ and $\tilde{\phi}$ satisfying
the following orthogonality conditions: \bea
&&\sum_{\a}\bar{\phi}^{(\a)}_{m,m-\e_i}(u)
~\phi^{(\a)}_{m,m-\e_j}(u)=\d_{ij},\label{Int1}\\[6pt]
&&\sum_{\a}\tilde{\phi}^{(\a)}_{m+\e_i,m}(u)
~\phi^{(\a)}_{m+\e_j,m}(u)=\d_{ij}.\label{Int2}\eea From these
conditions we  derive the ``completeness" relations:\bea
&&\sum_{k}\bar{\phi}^{(\a)}_{m,m-\e_k}(u)
~\phi^{(\b)}_{m,m-\e_k}(u)=\d_{\a\b},\label{Int3}\\[6pt]
&&\sum_{k}\tilde{\phi}^{(\a)}_{m+\e_{k},m}(u)
~\phi^{(\b)}_{m+\e_{k},m}(u)=\d_{\a\b}.\label{Int4}\eea

With the help of (\ref{Int1})-(\ref{Int4}), we  obtain, from the
face-vertex correspondence relation (\ref{Face-vertex}),
\begin{eqnarray}
 &&\left(\tilde{\phi}_{m+\e_{k},m}(u_1)\otimes
 {\rm id}\right)R_{12}(u_1-u_2) \left({\rm
id}\otimes\phi_{m+\e_{j},m}(u_2)\right)\no\\
 &&\qquad\quad= \sum_{i,l}W^{kl}_{ij}(u_1-u_2)\,
 \tilde{\phi}_{m+\e_{i}+\e_{j},m+\e_{j}}(u_1)\otimes
 \phi_{m+\e_{k}+\e_{l},m+\e_{k}}(u_2),\label{Face-vertex1}\\
 &&\left(\tilde{\phi}_{m+\e_{k},m}(u_1)\otimes
 \tilde{\phi}_{m+\e_{k}+\e_{l},m+\e_{k}}(u_2)\right)R_{12}(u_1-u_2)\no\\
 &&\qquad\quad= \sum_{i,j}W^{kl}_{ij}(u_1-u_2)\,
 \tilde{\phi}_{m+\e_{i}+\e_{j},m+\e_{j}}(u_1)\otimes
 \tilde{\phi}_{m+\e_{j},m}(u_2),\label{Face-vertex2}\\
 &&\left({\rm id}\otimes
 \bar{\phi}_{m,m-\e_{l}}(u_2)\right)R_{12}(u_1-u_2)
 \left(\phi_{m,m-\e_{i}}(u_1)\otimes {\rm id}\right)\no\\
 &&\qquad\quad= \sum_{k,j}W^{kl}_{ij}(u_1-u_2)\,
 \phi_{m-\e_{l},m-\e_{k}-\e_{l}}(u_1)\otimes
 \bar{\phi}_{m-\e_{i},m-\e_{i}-\e_{j}}(u_2),\label{Face-vertex3}\\
 &&\left(\bar{\phi}_{m-\e_{l},m-\e_{k}-\e_{l}}(u_1)\otimes
 \bar{\phi}_{m,m-\e_{l}}(u_2)\right)R_{12}(u_1-u_2)\no\\
 &&\qquad\quad= \sum_{i,j}W^{kl}_{ij}(u_1-u_2)\,
 \bar{\phi}_{m,m-\e_{i}}(u_1)\otimes
 \bar{\phi}_{m-\e_{i},m-\e_{i}
 -\e_{j}}(u_2).\label{Face-vertex4}
\end{eqnarray}

\section{Algebraic Bethe ansatz for the $A^{(1)}_{n-1}$ trigonometric vertex
model with non-diagonal open boundaries} \label{ABZ}
\setcounter{equation}{0} In this section, we shall demonstrate
that the intertwiners and the face-vertex correspondence relations
(\ref{Face-vertex})-(\ref{Face-vertex4})  play a fundamental role
in the construction of the eigenstates of the $A^{(1)}_{n-1}$
trigonometric vertex
 model with open boundary condition specified by
the K-matrices $K^{\pm}(u)$ given in
(\ref{K-matrix})-(\ref{Restriction}). In order to apply the
algebraic Bethe ansatz method, we need to transform the
fundamental exchange relation (\ref{Relation-Re}) from the  {\it
vertex picture\/} into its {\it face picture\/}  so that we can
construct the corresponding pseudo-vacuum and the associated Bethe
ansatz states.
\subsection{K-matrices in the ``face" picture}
Corresponding to the vertex type K-matrices (\ref{K-matrix}) and
(\ref{DK-matrix}), one introduces the following face type
K-matrices $\K$ and $\tilde{\K}$, as in \cite{Yan041}
\begin{eqnarray}
 &&\K(\l|u)^j_i=\sum_{\a,\b}\tilde{\phi}^{(\a)}
 _{\l-\e_i+\e_j,~\l-\e_i}(u)\,K^-(u)^{\a}_{\b}\,\phi^{(\b)}
 _{\l,~\l-\e_i}(-u),\label{K-F-1}\\
 &&\tilde{\K}(\l'|u)^j_i=\sum_{\a,\b}\bar{\phi}^{(\a)}
 _{\l',~\l'-\e_j}(-u)\,K^+(u)^{\a}_{\b}\,\phi^{(\b)}
 _{\l'-\e_j+\e_i,~\l'-\e_j}(u).\label{K-F-2}
\end{eqnarray}
Through straightforward calculations, we find the
$\l$($\l'$)-independent face type K-matrices have {\it diagonal\/}
forms\footnote{The spectral parameter $u$ and the boundary
parameter $\xi$ of the reduced double-row monodromy matrices
constructed from $\K(\l|u)$ will be shifted in each step of the
nested Bethe ansatz procedure \cite{Yan04}. Therefore, it is
convenient to specify the dependence on the boundary parameter
$\xi$ of $\K(\l|u)$, in terms of $k(u;\xi)_i$, in addition to the
spectral parameter $u$. }
\begin{eqnarray}
\K(\l|u)^j_i=\d_i^j\,k(u;\xi)_i,\quad
\tilde{\K}(\l'|u)^j_i=\d_i^j\,\tilde{k}(u)_i,~~i,j=1,\ldots,n,
\label{Diag-F}
\end{eqnarray}
where functions $\{k(u;\xi)_i=k(u)_i\}$ and $\{\tilde{k}(u)_i \}$
are respectively given by (\ref{K-matrix1}) and
(\ref{DK-matrix1}).

Although the K-matrices $K^{\pm}(u)$ given by (\ref{K-matrix}) and
(\ref{DK-matrix}) are generally non-diagonal (in the vertex
picture), after the face-vertex transformations (\ref{K-F-1}) and
(\ref{K-F-2}), the face type counterparts $\K(\l|u)$ and
$\tilde{\K}(\l'|u)$ become diagonal {\it simultaneously\/}. This
fact enables us  to apply the generalized algebraic Bethe ansatz
method developed in \cite{Yan04} for SOS type integrable models to
diagonalize the transfer matrix $\tau(u)$ (\ref{trans}).

\subsection{Exchange relations of double-row monodromy
matrix of face type}

By means of (\ref{Int3}), (\ref{Int4}), (\ref{K-F-2}) and
(\ref{Diag-F}), the transfer matrix $\t(u)$ (\ref{trans}) can be
recasted  into the following face type form:
\begin{eqnarray}
 \t(u)&=&tr(K^+(u)\mathbf{T}(u))\no\\
 &=&\sum_{\mu,\nu}tr\lt(\!K^+(u)\, \phi_{\l'-\e_{\mu}+\e_{\nu},
 \l'-\e_{\mu}}(u)\,\tilde{\phi}_{\l'-\e_{\mu}+\e_{\nu},
 \l'-\e_{\mu}}(u)~\mathbf{T}(u) \phi_{\l', \l'-\e_{\mu}}(-u)
 \bar{\phi}_{\l', \l'-\e_{\mu}}(-u)\rt)\no\\
 &=&\sum_{\mu,\nu}\bar{\phi}_{\l', \l'-\e_{\mu}}(-u)K^+(u)
 \,\phi_{\l'-\e_{\mu}+\e_{\nu},\l'-\e_{\mu}}(u)~
 \tilde{\phi}_{\l'-\e_{\mu}+\e_{\nu},
 \l'-\e_{\mu}}(u)~\mathbf{T}(u)\phi_{\l', \l'-\e_{\mu}}(-u)\no\\
 &=&\sum_{\mu,\nu}\tilde{\K}(\l'|u)_{\nu}^{\mu}\,\T(\l'|u)^{\nu}_{\mu}=
 \sum_{\mu}\tilde{k}(u)_{\mu}\,\T(\l'|u)^{\mu}_{\mu}.
 \label{De1}
\end{eqnarray}
Here we have introduced the face type double-row monodromy matrix
$\T(m|u)$,
\begin{eqnarray}
 \T(\l'|u)^{\nu}_{\mu}&=&\T(m|u)^{\nu}_{\mu}\lt|_{m=\l'}\rt.=
 \tilde{\phi}_{m-\e_{\mu}+\e_{\nu},
 m-\e_{\mu}}(u)~\mathbf{T}(u)\,\phi_{m,
 m-\e_{\mu}}(-u)\lt|_{m=\l'}\rt.\no\\
 &\equiv&
 \sum_{\a,\b}\tilde{\phi}^{(\b)}_{\l'-\e_{\mu}+\e_{\nu},
 \l'-\e_{\mu}}(u)~\mathbf{T}(u)^{\b}_{\a}\,\phi^{(\a)}_{\l',
 \l'-\e_{\mu}}(-u).\label{Mon-F}
\end{eqnarray}
Moreover from (\ref{Relation-Re}), (\ref{Face-vertex}) and
(\ref{Int4}) we derive the following exchange relations among
$\T(m|u)^{\nu}_{\mu}$ (see Appendix A for details):
\begin{eqnarray}
 &&\sum_{i_1,i_2}\sum_{j_1,j_2}~
 W^{i_0\,j_0}_{i_1\,j_1}(u_1-u_2)\,\T(m+\e_{j_1}+\e_{i_2}|u_1)
 ^{i_1}_{i_2}\no\\
 &&~~~~~~~~\times W^{j_1\,i_2}_{j_2\,i_3}(u_1+u_2)\,
 \T(m+\e_{j_3}+\e_{i_3}|u_2)^{j_2}_{j_3}\no\\[2pt]
 &&~~=\sum_{i_1,i_2}\sum_{j_1,j_2}~
 \T(m+\e_{j_1}+\e_{i_0}|u_2)
 ^{j_0}_{j_1}\,W^{i_0\,j_1}_{i_1\,j_2}(u_1+u_2)\no\\
 &&~~~~~~~~\times\T(m+\e_{j_2}+\e_{i_2}|u_1)^{i_1}_{i_2}\,
 W^{j_2\,i_2}_{j_3\,i_3}(u_1-u_2).\label{RE-F}
\end{eqnarray}

For convenience let us introduce the standard notation:
\begin{eqnarray}
\A(m|u)&=&\T(m|u)^1_1,~\B_i(m|u)= \T(m|u)^1_i,~\C_i(m|u)=
\T(m|u)^i_1,\quad i=2,\ldots,n,\label{Def-AB}\\
\D^j_i(m|u)&=&\T(m|u)^j_i-\d^j_iW^{j\,1}_{1\,j}(2u)\,\A(m|u),
\quad i,j=2,\ldots,n. \label{Def-D}
\end{eqnarray}
{}From (\ref{RE-F}), after some tedious calculation, we find the
commutation relations among $\A(m|u)$, $\D(m|u)$ and $\B(m|u)$
(see Appendix B for details). Here we give those which are
relevant for our purpose
\begin{eqnarray}
&&\A(m|u)\B_j(m+\e_{j}-\e_{1}|v)\no\\
&&\qquad=\frac{\sin(u+v)\sin(u-v-\eta)}{\sin(u+v+\eta)\sin(u-v)}
\,\B_j(m+\e_{j}-\e_{1}|v)\A(m+\e_{j}-\e_{1}|u)\no\\
&&\qquad\quad+\frac{\sin(\eta)\sin(2v)e^{i(u-v)}}{\sin(u-v)\sin(2v+\eta)}
\,\B_j(m+\e_{j}-\e_{1}|u)\A(m+\e_{j}-\e_{1}|v)\no\\
&&\qquad\quad-\frac{\sin(\eta)e^{i(u+v)}}{\sin(u+v+\eta)}\sum_{\a=2}^{n}
\,\B_{\a}(m+\e_{\a}-\e_{1}|u)\D^{\a}_j(m+\e_{j}-\e_{1}|v),\label{Rel-1}\\
&&\D^k_a(m|u)\B_j(m+\e_{j}-\e_{1}|v)\no\\
&&\qquad=
\frac{\sin(u-v+\eta)\sin(u+v+2\eta)}{\sin(u-v)\sin(u+v+\eta)}\no\\
&&\qquad\qquad\qquad\times\lt\{\sum_{\a_1,\a_2,\b_1,\b_2=2}^n
W^{k\,\,\,\,\b_2}_{\a_2\,\b_1} (u+v+\eta)
W^{\b_1\,\a_1}_{j\,\,\,\,a}(u-v)\rt.\no\\
&&\qquad\qquad\qquad\qquad\qquad\quad
\times\lt.\B_{\b_2}(m+\e_{k}+\e_{\b_2}-\e_{a}-\e_{1}|v)
\D^{\a_2}_{\a_1}(m+\e_{j}-\e_{1}|u)\rt\}\no\\
&&\qquad\quad-\frac{\sin(\eta)\sin(2u+2\eta)e^{-i(u-v)}}{\sin(u-v)\sin(2u+\eta)}\no\\
&&\qquad\qquad\qquad\times \lt\{
\sum_{\a,\b=2}^n\,W^{k\,\b}_{\a\,a}
(2u+\eta)\B_{\b}(m+\e_{\a}-\e_{1}|u)
\D^{\a}_{j}(m+\e_{j}-\e_{1})|v)\rt\}\no\\
&&\qquad\quad+\frac{\sin(\eta)\sin(2v)\sin(2u+2\eta)e^{-i(u+v)}}
{\sin(u+v+\eta)\sin(2v+\eta)\sin(2u+\eta)}\no\\[2pt]
&&\qquad\qquad\qquad\times\lt\{ \sum_{\a=2}^n
W^{k\,\a}_{j\,\,\,a}(2u+\eta)\, \B_{\a}(m+\e_{j}-\e_{1}|u)
\A(m+\e_{j}-\e_{1}|v)\rt\}, \label{Rel-2}\\[6pt]
&&\B_i(m+\e_{i}-\e_{1}|u)
\B_j(m+\e_{i}+\e_{j}-2\e_{1}|v)\no\\
&&\qquad=\sum_{\a,\b=2}^nW^{\b\,\a}_{j\,\,\,i} (u-v)
\,\B_{\b}(m+\e_{\b}-\e_{1}|v)
\,\B_{\a}(m+\e_{\a}+\e_{\b}-2\e_{1}|u).\label{Rel-3}
\end{eqnarray}

\subsection{Pseudo-Vacuum state}
The algebraic Bethe ansatz requires, in addition to the relevant
commutation relations (\ref{Rel-1})-(\ref{Rel-3}), a pseudo-vacuum
state  which is the common eigenstate of the operators $\A$,
$\D^i_i$ and is annihilated by the operators $\C_i$. In contrast
to the  models with {\it diagonal\/} $K^{\pm}(u)$
\cite{Skl88,Dev94}, for  models with {\it non-diagonal\/}
K-matrices, the usual highest-weight state
\begin{eqnarray}
 \lt(\begin{array}{l}1\\0\\\vdots\end{array}\rt)\otimes\cdots\otimes
 \lt(\begin{array}{l}1\\0\\\vdots\end{array}\rt),\label{O}
\end{eqnarray}
is no longer the pseudo-vacuum state. However, after the
face-vertex transformations (\ref{K-F-1}) and (\ref{K-F-2}), the
face type K-matrices $\K(\l|u)$ and $\tilde{\K}(\l|u)$ {\it
simultaneously\/} become diagonal. This suggests that one can
translate the $A^{(1)}_{n-1}$ trigonometric vertex model  with
non-diagonal K-matrices (\ref{K-matrix}) and (\ref{DK-matrix})
into the corresponding ``face" type model with {\it diagonal}
K-matrices $\K(\l|u)$ and $\tilde{\K}(\l|u)$ given by
(\ref{K-F-1}) and (\ref{K-F-2}) respectively. Then one can
construct the pseudo-vacuum in the ``face" picture and use the
generalized algebraic Bethe ansatz method \cite{Yan04} to
diagonalize the transfer matrix. Such a method has already been
successfully  used to diagonalize the transfer matrix of
$A^{(1)}_{1}$ trigonometric vertex model (XXZ model) with
non-diagonal K-matrices \cite{Yan043}. In this paper we shall
extend the  construction to the generic $A^{(1)}_{n-1}$ case.

Before introducing the pseudo-vacuum state, let us introduce  a
generic state in the quantum space by the  column vectors of the
intertwiner-matrix (\ref{In-matrix}) \bea |i_1,\ldots,i_N\rangle
_{m_0}^{m}&=&\phi_{m_0,m_0-\e_{i_N}}^N(-z_N)
\phi_{m_0-\e_{i_N},m_0-\e_{i_N}-\e_{i_{N-1}}}^{N-1}(-z_{N-1})\cdots\no\\
&&\qquad\times\phi_{m_0-\sum_{k=2}^N\e_{i_k},m_0-\sum_{k=1}^N\e_{i_k}}^1(-z_1),
\label{gstate} \eea where the vectors $m_0,m\in\Cb^n$ and
$m=m_0-\sum_{k=1}^N\e_{i_k}$,  the vector $\phi^k=id\otimes
id\cdots\otimes \stackrel{k-th}{\phi}\otimes id\cdots$.

Now let us evaluate the action of the face type monodromy  matrix
$\T(m|u)$ (\ref{Mon-F}) on the state (\ref{gstate}). Using the
definition of the vertex type double-row  monodromy matrix
${\mathbf{T}}(u)$ (\ref{Mon-V-1}) and relations
(\ref{Int3})-(\ref{Int4}), we can further write the face type
$\T(m|u)$ in the following form \bea
\T(m|u)^j_i&=&\tilde{\phi}_{m-\e_{i}+\e_{j},m-\e_{i}}(u)T(u)K^-(u)
T^{-1}(-u)\phi_{m,m-\e_{i}}(-u)\no\\
&\equiv&\T(m,m_0|u)^j_i\no\\
&=&\sum_{\mu,\nu}\tilde{\phi}_{m-\e_{i}+\e_{j},m-\e_{i}}(u)T(u)
\phi_{m_0-\e_{\nu}+\e_{\mu},m_0-\e_{\nu}}(u)
\tilde{\phi}_{m_0-\e_{\nu}+\e_{\mu},m_0-\e_{\nu}}(u)K^-(u)\no\\
&&\qquad\times \phi_{m_0,m_0-\e_{\nu}}(-u)
\bar{\phi}_{m_0,m_0-\e_{\nu}}(-u)T^{-1}(-u)\phi_{m,m-\e_{i}}(-u)\no\\
&=&\sum_{\mu,\nu}T(m-\e_{i},m_0-\e_{\nu}|u)^j_{\mu}
\K(m_0|u)^{\mu}_{\nu}S(m,m_0|u)^{\nu}_i.\label{Decomp} \eea Here
we have introduced  \bea
T(m,m_0|u)^j_{\mu}&=&\tilde{\phi}_{m+\e_{j},m}(u)
T(u)\phi_{m_0+\e_{\mu},m_0}(u),\label{Def-T}\\
S(m,m_0|u)^{\mu}_i&=&\bar{\phi}_{m_0,m_0-\e_{\mu}}(-u)
T^{-1}(-u)\phi_{m,m-\e_{i}}(-u),\label{Def-S}\\
\K(m_0|u)^j_{i}&=&\tilde{\phi}_{m_0-\e_{i}+\e_{j},m_0-\e_{i}}(u)
K^-(u)\phi_{m_0,m_0-\e_{i}}(-u). \label{F-V2} \eea We  can
evaluate the action of the operator $T(m,m_0|u)^j_{\mu}$ on the
state $|i_1,\ldots,i_N\rangle^m_{m_0}$ from the definition
(\ref{T-matrix}) and the  face-vertex correspondence relation
(\ref{Face-vertex}) \bea
&&T(m,m_0|u)^j_{\mu}|i_1,\ldots,i_N\rangle_{m_0}^m\no\\
&&\qquad=\tilde{\phi}^0_{m+\e_{j},m}(u)R_{01}(u+z_1)
\phi^1_{m_0-\sum_{k=2}^N\e_{i_k},m_0-\sum_{k=1}^N\e_{i_k}}(-z_1)\cdots\no\\
&&\qquad\qquad\times
R_{0N}(u+z_N)\phi^N_{m_0,m_0-\e_{i_N}}(-z_N)\phi^0_{m_0+\e_{\mu},m_0}(u)\no\\
&&\qquad
=\sum_{\b_1,i'_N}\tilde{\phi}^0_{m+\e_{j},m}(u)R_{01}(u+z_1)
\phi^1_{m_0-\sum_{k=2}^N\e_{i_k},m_0-\sum_{k=1}^N\e_{i_k}}(-z_1)\cdots\no\\
&&\qquad\qquad\times
R_{0\,N-1}(u+z_{N-1})\phi^0_{m_0+\e_{\mu}-\e_{i'_N},m_0-\e_{i_N}}(u)
\phi^{N-1}_{m_0-\e_{i_N},m_0-\e_{i_{N-1}}-\e_{i_N}}(-z_{N-1})\no\\
&&\qquad\qquad\times W^{\b_1\,i_N'}_{\,\mu\,\,i_N}(u+z_N)
\phi^N_{m_0+\e_{\mu},m_0+\e_{\mu}-\e_{i'_N}}(-z_N)\no\\
&&\qquad\vdots\no\\
&&\qquad=W^{j\,\,i'_1}_{\b_{N-1}\,i_1}(u+z_1)
W^{\b_{N-1}\,i'_2}_{\b_{N-2}\,i_2}(u+z_2)\cdots\no\\
&&\qquad\qquad\times W^{\b_1\,\,i'_N}_{\,\mu\,\,\,\,i_N}(u+z_N)\,
|i'_1,\ldots,i'_N\rangle_{m_0+\e_{\mu}}^{m+\e_{j}}.
\label{T-element}\eea In the last equation the repeated indices
imply summation over $1,2,\ldots n$. Noting the unitarity of the
R-matrix  (\ref{Unitarity}), $T^{-1}(-u)$ can be written \bea
T_0^{-1}(-u)=R_{N\,0}(u-z_N)\cdots R_{1\,0}(u-z_1).\eea Then by
the face-vertex correspondence relation (\ref{Face-vertex}) we can
evaluate the action of the operator $S(m,m_0|u)^{\mu}_i$ on the
state $|i_1,\ldots,i_N\rangle^m_{m_0}$,  similar to what we have
done for $T(m,m_0|u)$:  \bea
S(m,m_0|u)^{\mu}_i|i_1,\ldots,i_N\rangle_{m_0}^{m}&=&
W^{i'_N\,\mu}_{i_N\,\a_{N-1}}(u-z_N)
W^{i'_{N-1}\,\a_{N-1}}_{i_{N-1}\,\a_{N-2}}(u-z_{N-1})\cdots
\no\\
&&\qquad\times W^{i'_1\,\a_1}_{\,i_1\,\,\,i}(u-z_1)\,
|i'_1,\ldots,i'_N\rangle_{m_0-\e_{\mu}}^{m-\e_{i}}.
\label{S-element}\eea Here the repeated indices are summed over
$1,2,\ldots n$. Similarly, by the decomposition relation
(\ref{Decomp}) and the equations (\ref{T-element}),
(\ref{S-element}) we obtain the action of $\T(m|u)^j_i$ on the
state $|i_1,\ldots,i_N\rangle_{m_0}^{m}$:  \bea
&&\T(m|u)^j_i\,|i_1,\ldots,i_N\rangle_{m_0}^{m}\no\\
&&\qquad\equiv\T(m,m_0|u)^j_i\,|i_1,\ldots,i_N\rangle_{m_0}^{m}\no\\
&&\qquad=T(m-\e_{i},m_0-\e_{\nu}|u)^j_{\mu}
\K(m_0|u)^{\mu}_{\nu}S(m,m_0|u)^{\nu}_i
\,|i_1,\ldots,i_N\rangle_{m_0}^{m}\no\\
&&\qquad=W^{j\,\,i''_1}_{\b_{N-1}\,i'_1}(u+z_1)
W^{\b_{N-1}\,i''_2}_{\b_{N-2}\,i'_2}(u+z_2)
\cdots\no\\
&&\qquad\qquad\times W^{\b_1\,i''_N}_{\mu\,\,i'_N}(u+z_N)\,
\K(m_0|u)^{\mu}_{\nu}\, W^{i'_N\,\nu}_{i_N\,\a_{N-1}}(u-z_N)
\no\\
&&\qquad\qquad\times
W^{i'_{N-1}\,\a_{N-1}}_{i_{N-1}\,\a_{N-2}}(u-z_{N-1}) \cdots
W^{i'_1\,\a_1}_{\,i_1\,\,i}(u-z_1)\, |i''_1,\ldots,i''_N\rangle
_{m_0-\e_{\nu}+\e_{\mu}}^{m-\e_{i}+\e_{j}}. \label{Element-F}\eea
Here again it is understood that the repeated indices are summed
over $1,2,\ldots n$.

Specializing  the face-type parameters $\{(m_0)_i\}$ to the
boundary parameters $\{\l_i\}$, i.e. $m=\l$, in equation
(\ref{F-V2}), then from equation (\ref{Diag-F}) the corresponding
face type boundary K-matrix $\K(\l|u)$  becomes diagonal.  This
enables us to construct the pseudo-vacuum state of the model and
apply the algebraic Bethe ansatz method to diagonalize the {\it
double-row transfer matrices} (\ref{trans}) later.

Now, let us construct the pseudo-vacuum state $|\O\rangle$: \bea
|\O\rangle\equiv|vac\rangle^{\l-N\e_{1}}_{\l}=|1,\ldots,1\rangle
^{\l-N\e_{1}}_{\l},\label{Vac}\eea where $\l$ is related to the
boundary parameters $\{\l_i\}$ of the boundary K-matrix $K^{-}(u)$
in (\ref{K-matrix}). Then from equations (\ref{T-element}) and
(\ref{S-element}) we find that the actions of the operators
$T(\l-N\e_{1},\l|u)$ given by (\ref{Def-T}) and
$S(\l-N\e_{1},\l|u)$ given by (\ref{Def-S}) on the pseudo-vacuum
state (\ref{Vac}) are given by \bea
&&T(\l-N\e_{1},\l|u)^1_1\,|vac\rangle^{\l-N\e_{1}}_{\l}
=|vac\rangle^{\l-N\e_{1}+\e_1}_{\l+\e_{1}},\label{Action-1}\\
&&T(\l-N\e_{1},\l|u)^i_1\,|vac\rangle^{\l-N\e_{1}}_{\l}=0,
~~i=2,\ldots,n,\\
&&T(\l-N\e_{1},\l|u)^i_j\,|vac\rangle^{\l-N\e_{1}}_{\l}
=\d^i_j\prod_{k=1}^NW^{j1}_{\,j1}(u+z_k)\,
|vac\rangle^{\l-N\e_{1}+\e_{j}}_{\l+\e_{j}},\no\\
&&~~~~~~~~~~~~~~~~i,j=2,\ldots,n,\\
&&S(\l-N\e_{1},\l|u)^1_1\,|vac\rangle^{\l-N\e_{1}}_{\l}
=|vac\rangle^{\l-N\e_{1}-\e_{1}}_{\l-\e_{1}},\\
&&S(\l-N\e_{1},\l|u)^i_1\,|vac\rangle^{\l-N\e_{1}}_{\l}=0,
~~i=2,\ldots,n,\\
&&S(\l-N\e_{1},\l|u)^i_j\,|vac\rangle^{\l-N\e_{1}}_{\l}
=\d^i_j\prod_{k=1}^NW^{1j}_{\,1j}(u-z_k)\,
|vac\rangle^{\l-N\e_{1}-\e_{j}}_{\l-\e_{j}},\no\\
&&~~~~~~~~~~~~~~~~i,j=2,\ldots,n. \label{Action-2}\eea Noting that
the diagonal form of $\K(\l|u)$ (\ref{Diag-F}) and the above
equations, we derive \bea
&&\T(\l-N\e_{1},\l|u)^1_1\,|vac\rangle^{\l-N\e_{1}}_{\l}
=k(u;\xi)_1|vac\rangle^{\l-N\e_{1}}_{\l},\label{TF-V1}\\
&&\T(\l-N\e_{1},\l|u)^i_1\,|vac\rangle^{\l-N\e_{1}}_{\l}=0,
~~i=2,\ldots,n.\label{TF-V2} \eea Moreover, after a tedious
calculation, we have (see Appendix C for details)\bea
&&\T(\l-N\e_{1},\l|u)^i_j\,|vac\rangle^{\l-N\e_{1}}_{\l}\no\\
&&\qquad=\d^i_j \lt\{W^{j1}_{\,1j}(2u)k(u;\xi)_1 \mbox{{\Huge (}}1
-\prod_{k=1}^N W^{1j}_{\,1j}(u-z_k)
W^{j1}_{\,j1}(u+z_k)\rt)\no\\
&&\qquad\qquad\lt.+k(u;\xi)_j \prod_{k=1}^N W^{1j}_{\,1j}(u-z_k)
W^{j1}_{\,j1}(u+z_k)\,\rt\}|vac\rangle^{\l-N\e_{1}}_{\l},\no\\
&&\qquad\quad i,j=2,\ldots,n.\label{TF-V3} \eea

Keeping the definition of operators $\A$ (\ref{Def-AB}) and
$\D^i_j$ (\ref{Def-D}) in mind, and using the relations
(\ref{TF-V1})-(\ref{TF-V3}),   we find that the pseudo-vacuum
state given by (\ref{Vac}) satisfies the following equations \bea
\A(\l-N\e_{1}|u)\,|\O\rangle&=&k(u;\xi)_1
|\O\rangle,\label{A}\\
\D^a_j(\l-N\e_{1}|u)\,|\O\rangle&=&\d^a_j
\frac{\sin(2u)e^{i\eta}}{\sin(2u+\eta)}
k(u+\frac{\eta}{2};\xi-\frac{\eta}{2})_j\no\\
&&\qquad\times\lt\{\prod_{k=1}^N
\frac{\sin(u+z_k)\sin(u-z_k)}{\sin(u+z_k+\eta)\sin(u-z_k+\eta)}\rt\}
\,|\O\rangle,\no\\
a,j&=&2,\ldots,n,\label{D}\\
\C_i(\l-N\e_{1}|u)\,|\O\rangle&=&0,~i=2,\ldots,n,\\
\B_i(\l-N\e_{1}|u)\,|\O\rangle&\neq& 0,~i=2,\ldots,n, \eea as
required. In deriving the equation (\ref{D}), we have used the
following equation \bea k(u;\xi)_j-k(u;\xi)_1W^{j1}_{\,1j}(2u)=
\frac{\sin(2u)e^{i\eta}}{\sin(2u+\eta)}
k(u+\frac{\eta}{2};\xi-\frac{\eta}{2})_j. \eea

Therefore, we have constructed the pseudo-vacuum state
$|\O\rangle$ which is the common eigenstate of the operators $\A$,
$\D^{i}_{i},\,i=2,\ldots,n,$ and is annihilated  by the operators
$\C_i,\,i=2,\ldots,n.$ The operators $\B_i,\,i=2,\ldots,n,$ will
play the role of creation operators used to generate the Bethe
ansatz states.

\subsection{Nested Bethe ansatz}
Having derived the relevant commutation relations
(\ref{Rel-1})-(\ref{Rel-3}) and constructed the pseudo-vacuum
state (\ref{Vac}), we now apply the generalized algebraic Bethe
ansatz method  developed in \cite{Yan04} to solve the eigenvalue
problem for the transfer matrices (\ref{trans}) of the
$A^{(1)}_{n-1}$ trigonometric vertex model with open boundary
condition specified by the K-matrices $K^{\pm}(u)$ given in
(\ref{K-matrix})-(\ref{Restriction}).

For convenience, let us introduce a set of non-negative  integers
$\{N_i|i=1,\ldots,n-1\}$ with  $N_1=N$ and complex parameters
$\{v^{(i)}_k|~k=1,2,\ldots,N_{i+1},~i=0,1,\ldots,n-2\}$. As in the
usual nested Bethe ansatz method
\cite{Bab82,Sch83,Dev94,Hou03,Yan04,Yan044}, the parameters
$\{v^{(i)}_k\}$ will be used to specify the eigenvectors of the
corresponding reduced transfer matrices.  They will be constrained
later by the Bethe ansatz equations. For convenience, we adopt the
following convention:
\begin{eqnarray}
 v_k=v^{(0)}_k,~k=1,2,\ldots, N.
\end{eqnarray}
We will seek the common eigenvectors (i.e. the so-called Bethe
states) of the transfer matrix in the form
\begin{eqnarray}
&&|v_1,\ldots,v_{N}\rangle =\sum_{i_1,\ldots,i_{N}=2}^n
F^{i_1,i_2,\ldots,i_{N}}\,
\B_{i_1}(\l'+\e_{i_1}-\e_{1}|v_{1})\,\B_{i_2}
(\l'+\e_{i_1}+\e_{i_2}-2\e_{1}|v_2)\cdots\no\\
&&\qquad\qquad\qquad\qquad\qquad\times \B_{i_{N-1}}
(\l'+\sum_{k=1}^{N-1}\e_{i_k}-(N-1)\e_{1}|v_{N-1})\no\\
&&\qquad\qquad\qquad\qquad\qquad\times\B_{i_{N}}
(\l'+\sum_{k=1}^{N}\e_{i_k}-N\e_{1}|v_{N})\,
|\O\rangle.\label{Eigenstate1}
\end{eqnarray}
The summing indices in the above equation should obey the
following restriction: \bea
\l'+\sum_{k=1}^N\e_{i_k}=\l,\label{Restriction1}\eea where $\l'$
and $\l$ are the boundary parameters which satisfy the restriction
(\ref{Restriction}). The condition (\ref{Restriction1}) leads to
\begin{eqnarray}
&&|v_1,\ldots,v_{N}\rangle =\sum_{i_1,\ldots,i_{N}=2}^n
F^{i_1,i_2,\ldots,i_{N}}\,
\B_{i_1}(\l'+\e_{i_1}-\e_{1}|v_{1})\,\B_{i_2}
(\l'+\e_{i_1}+\e_{i_2}-2\e_{1}|v_2)\cdots\no\\
&&\qquad\qquad\qquad\qquad\qquad\times \B_{i_{N-1}}
(\l'+\sum_{k=1}^{N-1}\e_{i_k}-(N-1)\e_{1}|v_{N-1})\no\\
&& \qquad\qquad\qquad\qquad\qquad\times \B_{i_{N}}
(\l-N\e_{1}|v_{N})\,|\O\rangle.\label{Eigenstate}
\end{eqnarray}

With the help of (\ref{De1}), (\ref{Def-AB}) and (\ref{Def-D}) we
rewrite the transfer matrix (\ref{trans}) in terms of the
operators $\A$ and $\D^i_i$
\begin{eqnarray}
 \t(u)&=&\sum_{\nu=1}^n\tilde{k}(u)_{\mu}\,\T(\l'|u)^{\mu}_{\mu}
 \no\\
 &=&\tilde{k}(u)_1\,\A(\l'|u)+\sum_{i=2}^n\tilde{k}(u)_i
 \,\T(\l'|u)^i_i \no\\
 &=&\tilde{k}(u)_1\,\A(\l'|u)+ \sum_{i=2}^n\tilde{k}(u)_i
 \,W^{i1}_{1\,i}(2u)\,\A(\l'|u)\no\\
 &&\phantom{\tilde{k}(u)_1\A(\l'|u)}  + \sum_{i=2}^n\tilde{k}(u)_i
 \left(\vTm\T(\l'|u)^i_i- W^{i1}_{1\,i}(2u)\A(\l'|u)\right)\no\\
 &=& \sum_{i=1}^n\tilde{k}(u)_i
 W^{i1}_{1\,i}(2u)\,\A(\l'|u)\no\\
 &&~~~~~~~~+ \sum_{i=2}^n\tilde{k}^{(1)}(u+\frac{\eta}{2})_i
 \left(\vTm\T(\l'|u)^i_i-
 W^{i1}_{1\,i}(2u)\A(\l'|u)\right)\no\\
 &=&\a^{(1)}(u)\,\A(\l'|u)+\sum_{i=2}^n
 \tilde{k}^{(1)}(u+\frac{\eta}{2})_i\,\D(\l'|u)^i_i.
 \label{trans1}
\end{eqnarray}
Here we have used (\ref{Def-D}) and introduced the function
$\a^{(1)}(u)$,
\begin{eqnarray}
 \a^{(1)}(u)=\sum_{i=1}^n\tilde{k}(u)_i\,
 W^{i1}_{1\,i}(2u),\label{function-a}
\end{eqnarray}
and the reduced
 K-matrix $\tilde{\K}^{(1)}(\l'|u)$ with the elements given by
\begin{eqnarray}
\tilde{\K}^{(1)}(\l'|u)^j_i&=&
 \d^j_i\,\tilde{k}^{(1)}(u)_i,\qquad\qquad\qquad
 i,j=2,\ldots,n,\label{Reduced-K1}\\
\tilde{k}^{(1)}(u)_i&=& \tilde{k}(u-\frac{\eta}{2})_i,
\qquad\qquad\qquad
 i=2,\ldots,n.
 \label{Reduced-K2}
\end{eqnarray}
To carry out the nested Bethe ansatz process
\cite{Dev94,Yan04,Yan044} for  the $A^{(1)}_{n-1}$ type models
with the open boundary conditions, one needs to introduce a set of
reduced K-matrices $\{\tilde{\K}^{(b)}(\l'|u)|b=0,\ldots,n-1\}$
\cite{Yan04} which include the original one
$\tilde{\K}(\l'|u)=\tilde{\K}^{(0)}(\l'|u)$ and the ones in
(\ref{Reduced-K1}) and (\ref{Reduced-K2}):
\begin{eqnarray}
 \tilde{\K}^{(b)}(\l'|u)^j_i&=&
 \d^j_i\,\tilde{k}^{(b)}(u)_i,\quad i,j=b+1,\ldots,n,
 \quad b=0,\ldots,n-1,
 \label{Reduced-K3}\\
 \tilde{k}^{(b)}(u)_i&=&
\tilde{k}(u-b\frac{\eta}{2})_i,\quad i=b+1,\ldots,n,
 \quad b=0,\ldots,n-1.
 \label{Reduced-K4}
\end{eqnarray}
Moreover we introduce a set of functions
$\{\a^{(b)}(u)|b=1,\ldots,n-1\}$ (including the one in
(\ref{function-a})) related to the reduced K-matrices
$\tilde{\K}^{(b)}(\l'|u)$
\begin{eqnarray}
 \a^{(b)}(u)=\sum_{i=b}^{n}W^{ib}_{\,bi}(2u)\,
 \tilde{k}^{(b-1)}(u)_i,\quad b=1,\ldots,n.\label{function-a-1}
\end{eqnarray}

Carrying out the nested Bethe ansatz, we finally find that, with
the coefficients $F^{i_1,i_2,\cdots,i_{N}}$ in (\ref{Eigenstate})
properly chosen, the Bethe state $|v_1,\ldots,v_{N}\rangle $ is
the eigenstate of the transfer matrix (\ref{trans}),
\begin{eqnarray}
 \t(u)\,|v_1,\ldots,v_{N}\rangle=\L(u;\xi,\{v_k\})\,
 |v_1,\ldots,v_{N}\rangle,\end{eqnarray} with eigenvalue given by
 \begin{eqnarray}
 &&\L(u;\xi,\{v_k\})\no\\[4pt]
 &&\qquad=\a^{(1)}(u)k(u;\xi)_1\,
 \prod_{k=1}^{N}\frac{\sin(u+v_k)\sin(u-v_k-\eta)}
 {\sin(u+v_k+\eta)\sin(u-v_k)}\no\\[4pt]
 &&\qquad\quad+\frac{\sin(2u)e^{i\eta}} {\sin(2u+\eta)}\,
\lt\{ \prod_{k=1}^{N}\frac{\sin(u-v_k+\eta)\s(u+v_k+2\eta)}
 {\sin(u-v_k)\sin(u+v_k+\eta)}\rt.\no\\[4pt]
 &&\qquad\qquad\qquad
 \times \prod_{k=1}^{N}\frac{\sin(u+z_k)\sin(u-z_k)}
 {\sin(u+z_k+\eta)\sin(u-z_k+\eta)}\no\\[4pt]
&&\qquad\qquad\qquad\times
\lt.\L^{(1)}(u+\frac{\eta}{2};\xi-\!\frac{\eta}{2},\{v^{(1)}_k\})\rt\}.
\label{Eigenvalue1}
\end{eqnarray}
The eigenvalues
$\{\L^{(j)}(u;\xi,\{v^{(j)}_{k}\})|j=0,\ldots,n-1\}$ (with
$\L(u;\xi,\{v_{k}\})=\L^{(0)}(u;\xi,\{v^{(0)}_{k}\})$) of the
reduced transfer matrices are given by the following recurrence
relations:
\begin{eqnarray}
&& \L^{(j)}(u;\xi^{(j)},\{v^{(j)}_k\})\no\\[4pt]
&&\qquad=\a^{(j+1)}(u)
 k(u;\xi^{(j)})_{j+1}\,
 \prod_{k=1}^{N_{j+1}}\frac{\sin(u+v^{(j)}_k)\sin(u-v^{(j)}_k-\eta)}
 {\sin(u+v^{(j)}_k+\eta)\sin(u-v^{(j)}_k)}\no\\[4pt]
 &&\qquad\quad+\frac{\sin(2u)e^{i\eta}}
 {\sin(2u+\eta)}\,
 \lt\{\prod_{k=1}^{N_{j+1}}\frac{\sin(u-v^{(j)}_k+\eta)\sin(u+v^{(j)}_k+2\eta)}
 {\sin(u-v^{(j)}_k)\sin(u+v^{(j)}_k+\eta)}\rt.\no\\[4pt]
 &&\qquad\qquad\qquad\times
 \prod_{k=1}^{N_j}\frac{\sin(u+z^{(j)}_k)\sin(u-z^{(j)}_k)}
 {\sin(u+z^{(j)}_k+\eta)\sin(u-z^{(j)}_k+\eta)}\no\\[4pt]
&&\qquad\qquad\qquad\times
\lt.\L^{(j+1)}(u+\frac{\eta}{2};\xi^{(j)}-
\!\frac{\eta}{2},\{v^{(j+1)}_k\})\rt\},\no\\
&&\qquad\qquad j=1,\ldots,n-2, \label{Eigenvalue2}\\[6pt]
&&\L^{(n-1)}(u;\xi^{(n-1)})=\tilde{k}^{(n-1)}(u)_n\,
k(u;\xi^{(n-1)})_n.\label{Eigenvalue3}
\end{eqnarray}
The reduced boundary parameters $\{\xi^{(j)}\}$ and inhomogeneous
parameters $\{z^{(j)}_k\}$ are given by
\begin{eqnarray}
 \xi^{(j+1)}=\xi^{(j)}-\frac{\eta}{2},
 \qquad z^{(j+1)}_k=v^{(j)}_k+\frac{\eta}{2},
 \qquad j=0,\ldots,n-2.\label{Parameters}
\end{eqnarray}
Here we have adopted the convention: $\xi=\xi^{(0)}$,
$z^{(0)}_k=z_k$. The complex parameters  $\{v^{(j)}_k\}$ satisfy
the following Bethe ansatz equations:
\begin{eqnarray}
 &&\a^{(1)}(v_s)k(v_s;\xi)_1\frac{\sin(2v_s+\eta)e^{-i\eta}}
 {\sin(2v_s+2\eta)}\no\\[4pt]
 &&\qquad\qquad\qquad\qquad\times \prod_{k\ne s,k=1}^{N_1}
 \frac{\sin(v_s+v_k)\sin(v_s-v_k-\eta)} {\sin(v_s+v_k+2\eta)
 \sin(v_s-v_k+\eta)}\no\\[4pt]
 &&\qquad = \prod_{k=1}^{N} \frac{\sin(v_s+z_k)\sin(v_s-z_k)}
 {\sin(v_s+z_k+\eta)\sin(v_s-z_k+\eta)}\no\\[4pt]
 &&\qquad\qquad\qquad\qquad
 \times\L^{(1)}(v_s+\frac{\eta}{2};\xi-\!\frac{\eta}{2},\{v^{(1)}_k\}),
 \label{BA1}\\[6pt]
 &&\a^{(j+1)}(v^{(j)}_s)k(v^{(j)}_s;\xi^{(j)})_{j+1}\,
 \frac{\sin(2v^{(j)}_s+\eta)e^{-i\eta}}
 {\sin(2v^{(j)}_s+2\eta)}\no\\[4pt]
&&\qquad\qquad\qquad\qquad\times \prod_{k\ne s,k=1}^{N_{j+1}}
 \frac{\sin(v^{(j)}_s+v^{(j)}_k)\sin(v^{(j)}_s-v^{(j)}_k-\eta)}
 {\sin(v^{(j)}_s+v^{(j)}_k+2\eta)\sin(v^{(j)}_s-v^{(j)}_k+\eta)}\no\\[4pt]
 &&\qquad = \prod_{k=1}^{N_j}
 \frac{\sin(v^{(j)}_s+z^{(j)}_k)\sin(v^{(j)}_s-z^{(j)}_k)}
 {\sin(v^{(j)}_s+z^{(j)}_k+\eta)\sin(v^{(j)}_s-z^{(j)}_k+\eta)}\no\\[4pt]
 &&\qquad\qquad\qquad\qquad
 \times\L^{(j+1)}(v^{(j)}_s+\frac{\eta}{2};
 \xi^{(j)}-\!\frac{\eta}{2},\{v^{(j+1)}_k\})
 ,\no\\
 &&\qquad\qquad\qquad j=1,\ldots,n-2.
 \label{BA2}
\end{eqnarray}

\section{Conclusions}
\label{Con} \setcounter{equation}{0}

We have studied the $A^{(1)}_{n-1}$  trigonometric vertex model
with integrable open boundary condition described by the {\it
generic non-diagonal\/} boundary K-matrix $K^-(u)$ given in
(\ref{K-matrix}) and its dual $K^+(u)$ given in (\ref{DK-matrix})
with restriction (\ref{Restriction}). In addition to the two {\it
discrete\/} (positive integers) parameters $l$ and $l'$, the total
number of the {\it independent\/} free boundary parameters:
$\xi,\,\bar\xi$, $\rho$ and $\l_i,\,i=1,\ldots,n$, is actually
$n+3$. Although the K-matrices given in (\ref{K-matrix}) and
(\ref{DK-matrix}) are {\it non-diagonal\/} in the vertex picture,
they  become diagonal {\it simultaneously\/} in the ``face"
picture after the face-vertex transformation  given by
(\ref{K-F-1})-(\ref{Diag-F}). This  fact {\it enables\/} us to
successfully construct the corresponding pseudo-vacuum state
$|\O\rangle$ (\ref{Vac}) and apply the algebraic Bethe ansatz
method to diagonalize the corresponding {\it double-row transfer
matrices\/}.  The eigenvalues of the transfer matrices and
associated Bethe ansatz equations are given by
(\ref{Eigenvalue1}), (\ref{Eigenvalue2})-(\ref{Parameters}), and
(\ref{BA1}), (\ref{BA2}). Taking the rational limit of our results
with $\{\lim_{\eta\rightarrow 0}(\eta\l_i)\}$ being kept finite,
we recover the results obtained in \cite{Gal04}.

\section*{Acknowledgements}
This work was financially supported by the Australian Research
Council.

\section*{Appendix A: The exchange relation of
$\T$} \setcounter{equation}{0}
\renewcommand{\theequation}{A.\arabic{equation}}

The starting point for deriving the exchange relations
(\ref{RE-F}) among $\T(m|u)^{\nu}_{\mu}$ is the exchange relation
(\ref{Relation-Re}). Multiplying both sides of (\ref{Relation-Re})
from the right by $\phi_{m+\e_{i_3},\,m}(-u_1) \otimes
\phi_{m+\e_{i_3}+\e_{j_3},\,m+\e_{i_3}}(-u_2)$, and using the
face-vertex correspondence relation (\ref{Face-vertex}) and the
``completeness" relation (\ref{Int4}), we have, for the L.H.S. of
the resulting relation, \bea
{\rm L.H.S.}&=&R_{12}(u_1-u_2)\mathbf{T}_1(u_1)R_{21}(u_1+u_2)\no\\
&&\qquad\quad\times (\phi_{m+\e_{i_3},\,m}(-u_1)\otimes
\mathbf{T}(u_2)\,
\phi_{m+\e_{i_3}+\e_{j_3},\,m+\e_{i_3}}(-u_2))\no\\
&=&R_{12}(u_1-u_2)\mathbf{T}_1(u_1)R_{21}(u_1+u_2)
(\phi_{m+\e_{i_3},\,m}(-u_1)\otimes 1)\no\\
&&\qquad\quad\times(1\otimes \{\sum_{j_2}
\phi_{m+\e_{i_3}+\e_{j_2},\,m+\e_{i_3}}(u_2)
\tilde{\phi}_{m+\e_{i_3}+\e_{j_2},\,m+\e_{i_3}}(u_2)\no\\
&&\qquad\quad\times \lt.\mathbf{T}(u_2)
\phi_{m+\e_{i_3}+\e_{j_3},\,m+\e_{i_3}}(-u_2)\rt\})\no\\
&=&\sum_{j_2}R_{12}(u_1-u_2)\mathbf{T}_1(u_1)R_{21}(u_1+u_2)\no\\
&&\qquad\quad\times\lt(\phi_{m+\e_{i_3},\,m}(-u_1)\otimes
\phi_{m+\e_{i_3}+\e_{j_2},\,m+\e_{i_3}}(u_2)\rt)
\T(m+\e_{i_3}+\e_{j_3}|u_2)^{j_2}_{j_3}
\no\\
&=&\sum_{i_2}\sum_{j_1,j_2}R_{12}(u_1-u_2)\mathbf{T}_1(u_1)
W^{j_1\,i_2}_{j_2\,i_3}(u_1+u_2)\no\\
&&\qquad\quad\times
\lt(\phi_{m+\e_{i_3}+\e_{j_2},\,m+\e_{j_1}}(-u_1)\otimes
\phi_{m+\e_{j_1},\,m}(u_2)\rt)\T(m+\e_{i_3}+\e_{j_3}|u_2)^{j_2}_{j_3}
\no\\
&&\vdots\no\\
&=&\sum_{i_0,j_0} (\phi_{m+\e_{i_0},\,m}(u_1)\otimes
\phi_{m+\e_{i_0}+\e_{j_0},\,m+\e_{i_0}}(u_2))\no\\
&&\qquad\quad\times \lt\{\sum_{i_1,i_2}
\sum_{j_1,j_2}W^{i_0\,j_0}_{i_1\,j_1}(u_1-u_2)
\T(m+\e_{i_2}+\e_{j_1}|u_1)^{i_1}_{i_2}\rt.\no\\
&&\qquad\quad\qquad\quad\times
\lt.W^{j_1\,i_2}_{j_2\,i_3}(u_1+u_2)
\T(m+\e_{i_3}+\e_{j_3}|u_2)^{j_2}_{j_3}\rt\}.\label{LHS} \eea
Similarly for the R.H.S. of the resulting relation, we obtain \bea
{\rm R.H.S.}&=& \sum_{i_0,j_0}
\lt(\phi_{m+\e_{i_0},\,m}(u_1)\otimes
\phi_{m+\e_{i_0}+\e_{j_0},\,m+\e_{i_0}}(u_2)\rt)\no\\
&&\qquad\quad\times\lt\{\sum_{i_1,i_2}
\sum_{j_1,j_2}\T(m+\e_{i_0}+\e_{j_1}|u_2)^{j_0}_{j_1}
W^{i_0\,j_1}_{i_1\,j_2}(u_1+u_2)\rt.
\no\\
&&\qquad\quad\qquad\quad\times
\lt.\T(m+\e_{i_2}+\e_{j_2}|u_1)^{i_1}_{i_2}
W^{j_2\,i_2}_{j_3\,i_3}(u_1-u_2)\rt\}.\label{RHS} \eea Note that
intertwiners are linearly independent, which follows from
(\ref{Det}). Thus  we obtain the exchange relation (\ref{RE-F}) by
comparing (\ref{LHS}) with (\ref{RHS}).

\section*{Appendix B: The relevant commutation relations}
\setcounter{equation}{0}
\renewcommand{\theequation}{B.\arabic{equation}}

Let us introduce  \bea
&&D^j_i(m|u)=\T(m|u)^j_i,~i,j=2,\ldots,n.\eea The starting point
for deriving the commutation relations among
$\A(m|u),~\D^j_i(m|u)$ and $\B_i(m|u)$ $(i,j=2,\ldots,n)$ is the
exchange relation (\ref{RE-F}).

For $i_0=j_0=j_3=1,~ i_3=i\neq 1$, we obtain \bea
&&\A(m+2\e_{1}|v)\B_j(m+\e_{j}+\e_{1}|u)\no\\
&&\qquad=\frac{\sin(u+v)\sin(u-v+\eta)}{\sin(u+v+\eta)\sin(u-v)}
\B_j(m+\e_{j}+\e_{1}|u)\A(m+\e_{j}+\e_{1}|v)\no\\
&&\qquad\quad-\frac{\sin(\eta)\sin(u+v)e^{-i(u-v)}}
{\sin(u-v)\sin(u+v+\eta)}
\B_j(m+\e_{j}+\e_{1}|v)\A(m+\e_{j}+\e_{1}|u)\no\\
&&\qquad\quad-\frac{\sin(\eta)e^{i(u+v)}} {\sin(u+v+\eta)}
\sum_{\a=2}^{n}
\B_{\a}(m+\e_{\a}+\e_{1}|v)D^{\a}_j(m+\e_{j}+\e_{1}|u).\label{AB}
\eea The commutation relation (\ref{Rel-1}) is a simple
consequence of (\ref{AB}), (\ref{Def-AB}) and (\ref{Def-D}).

\vspace{0.5truecm}

For $i_0=k\neq 1$, $j_0=1$, $i_3=i\neq 1$ and $j_3=j\neq 1$, we
obtain \bea
&&D^k_a(m+\e_{a}+\e_{1}|u)\B_j(m+\e_{j}+\e_{a}|v)\no\\
&&\qquad=\sum_{\a_1,\a_2,\b_1,\b_2=2}^n
\frac{W^{k\,\,\,\b_2}_{\a_2\,\b_1}(u+v)W^{\b_1\,\a_1}_{j\,\,\,\,a}(u-v)}
{W^{k1}_{\,k1}(u-v)W^{1a}_{\,1a}(u+v)}\no\\
&&\qquad\qquad\qquad\qquad\qquad\qquad\quad\times
\B_{\b_2}(m+\e_{k}+\e_{\b_2}|v)
D^{\a_2}_{\a_1}(m+\e_{a}+\e_{j}|u)\no\\
&&\qquad\quad+\sum_{\a=2}^n
\frac{W^{k\,1}_{1\,k}(u+v)W^{k\,\a}_{j\,\,\,a}(u-v)}
{W^{k1}_{\,k1}(u-v)W^{1a}_{\,1a}(u+v)} \A(m+\e_{k}+\e_{1}|v)
\B_{\a}(m+\e_{a}+\e_{j}|u)\no\\
&&\qquad\quad-\sum_{\a,\b=2}^n
\frac{W^{k\,1}_{1\,k}(u-v)W^{k\,\a}_{\b\,a}(u+v)}
{W^{k1}_{\,k1}(u-v)W^{1a}_{\,1a}(u+v)} \B_{\a}(m+\e_{k}+\e_{\a}|u)
D^{\b}_{j}(m+\e_{a}+\e_{j}|v)\no\\
&&\qquad\quad- \frac{W^{k\,1}_{1\,k}(u-v)W^{k\,1}_{1\,a}(u+v)}
{W^{k1}_{\,k1}(u-v)W^{1a}_{\,1a}(u+v)} \A(m+\e_{k}+\e_{1}|u)
\B_{j}(m+\e_{a}+\e_{j}|v). \label{DB}\eea In order to separate the
contribution of $\A$ and $D^j_a$ in the above relations, one needs
to introduce the operator $\D^j_a$ as (\ref{Def-D})
(cf.\cite{Skl88}). Then we can derive the commutation relations
 among $\D^k_a$ and $\B_j$ from (\ref{DB}) \bea
&&\D^k_a(m|u)\B_j(m+\e_{j}-\e_{1}|v)\no\\
&&\qquad=\sum_{\a_1,\a_2,\b_1,\b_2=2}^n
\frac{W^{k\,\,\,\b_2}_{\a_2\,\b_1}(u+v)
W^{\b_1\,\a_1}_{j\,\,\,\,a}(u-v)} {W^{k1}_{\,k1}(u-v)
W^{1a}_{\,1a}(u+v)}\no\\
&&\qquad\qquad\qquad\qquad\qquad\qquad\quad\times
\B_{\b_2}(m+\e_{k}+\e_{\b_2}-\e_{a}-\e_{1}|v)
\D^{\a_2}_{\a_1}(m+\e_{j}-\e_{1}|u)\no\\
&&\qquad\quad-\sum_{\a,\b=2}^n \frac{W^{k\,1}_{1\,k}(u-v)
W^{k\,\a}_{\b\,a}(u+v)} {W^{k1}_{\,k1}(u-v)
W^{1a}_{\,1a}(u+v)}\no\\
&&\qquad\qquad\qquad\qquad\qquad\qquad\quad\times
\B_{\a}(m+\e_{\b}-\e_{1}|u)
\D^{\b}_{j}(m+\e_{j}-\e_{1}|v)\no\\
&&\qquad\quad+\sum_{\a,\b_1,\b_2=2}^n
\frac{W^{k\,\,\b_2}_{\a\,\b_1}(u+v) W^{\b_1\,\a}_{j\,\,a}(u-v)}
{W^{k1}_{\,k1}(u-v) W^{1a}_{\,1a}(u+v)}
W^{\a 1}_{1\,\a}(2u)\no\\
&&\qquad\qquad\qquad\qquad\qquad\qquad\quad\times
\B_{\b_2}(m+\e_{\b_2}+\e_{k}-\e_{a}-\e_{1}|v)
\A(m+\e_{j}-\e_{1}|u)\no\\
&&\qquad\quad-\sum_{\a=2}^n \frac{W^{k\,1}_{1\,k}(u-v)
W^{k\,\a}_{j\,\,a}(u+v)} {W^{k1}_{\,k1}(u-v) W^{1a}_{\,1a}(u+v)}
W^{j1}_{\,1j}(2v)\no\\
&&\qquad\qquad\qquad\qquad\qquad\qquad\quad\times
\B_{\a}(m+\e_{\a}+\e_{k}-\e_{a}-\e_{1}|u)
\A(m+\e_{j}-\e_{1}|v)\no\\
&&\qquad\quad+\sum_{\a=2}^n \frac{W^{k\,1}_{1\,k}(u+v)
W^{k\,\a}_{j\,a}(u-v)} {W^{k1}_{\,k1}(u-v) W^{1a}_{\,1a}(u+v)}
\A(m+\e_{k}-\e_{a}|v)\B_{\a}(m+\e_{j}-\e_{1}|u)\no\\
&&\qquad\quad-\frac{\sin(u-v+\eta)\sin(u+v+\eta)\sin(\eta)e^{-2iu}}
{\sin(u-v)\sin(u+v)\sin(2u+\eta)}\no\\
&&\qquad\qquad\qquad\qquad\qquad\qquad\quad\times \d^k_a
\A(m-\e_{a}+\e_{k}|u)\B_j(m+\e_{j}-\e_{1}|v). \eea We have used
the following equation \bea &&\frac{W^{k1}_{1\,k}(u-v)
W^{k1}_{1\,k}(u+v)} {W^{k1}_{\,k1}(u-v) W^{1a}_{\,1a}(u+v)}
+W^{k1}_{1\,k}(2u)\no\\
&&\qquad=\frac{\sin(u-v+\eta)\sin(u+v+\eta)\sin(\eta)e^{-2iu}}
{\sin(u-v)\sin(u+v)\sin(2u+\eta)},\label{B-6} \eea to derive the
last term on the right side of the above equation. After some long
tedious calculation, we finally obtain the commutation relation
(\ref{Rel-2}) from (\ref{B-6}) by noting the definitions
(\ref{Def-AB}) and (\ref{Def-D}).

\vspace{0.5truecm}

For $i_0=j_0=1$, $i_3=i\neq 1$ and $j_3=j\neq 1$, we obtain \bea
&&\B_i(m+\e_{i}+\e_{1}|u) \B_j(m+\e_{i}+\e_{j}|v)\no\\
&&\qquad= \sum_{\b,\a=2}^n\frac{W^{\b\a}_{j\,i}(u-v)
W^{1\b}_{\,1\b}(u+v)} {W^{11}_{\,11}(u-v)
W^{1i}_{\,1i}(u+v)}\B_{\b}(m+\e_{\b}+\e_{1}|v)
\B_{\a}(m+\e_{i}+\e_{j}|u),\eea which leads to the commutation
relation (\ref{Rel-3}).

\section*{Appendix C: The action of $\T^i_j$ on the pseudo-vacuum state}
\setcounter{equation}{0}
\renewcommand{\theequation}{C.\arabic{equation}}

Carrying out the calculation similar to that leading to
(\ref{TF-V1}) and (\ref{TF-V2}), we have \bea
&&\T(\l-N\e_{1},\l|u)^i_j\,|vac\rangle^{\l-N\e_{1}}_{\l}\no\\
&&\qquad= k(u;\xi)_1T(\l-N\e_{1}-\e_{j},\l-\e_{1}|u)^i_1
S(\l-N\e_{1},\l|u)^1_j\,
|vac\rangle^{\l-N\e_{1}}_{\l}\no\\
&&\qquad\qquad+\d^i_jk(u;\xi)_j\prod_{k=1}^N W^{1j}_{\,1j}(u-z_k)
W^{j1}_{\,j1}(u+z_k)\no\\
&&\qquad\qquad\qquad\qquad\qquad\times
|vac\rangle^{\l-N\e_{1}}_{\l},~~i,j=2,\ldots,n.\label{C-1}\eea The
first term on the right hand of the above equation is obtained as
follows.

{}From ``RLL" relation (\ref{Relation1}), we  derive the following
exchange relations  \bea
T_1(u)R_{12}(2u)T^{-1}_2(-u)=T^{-1}_2(-u)R_{12}(2u)T_1(u). \eea
Multiplying both sides of the  above equation from the left by
$\tilde{\phi}_{\l-N\e_{1}-\e_{i}+\e_{j},\l-N\e_{1}-\e_{i}}(u)
\otimes\bar{\phi}_{\l+\e_{1},\l}(-u)$ and from the right by
$\phi_{\l+\e_{1},\l}(u)\otimes
\phi_{\l-N\e_{1},\l-N\e_{1}-\e_{i}}(-u)$, we obtain  the following
exchange relation from the face-vertex correspondence relations
(\ref{Face-vertex}) and (\ref{Face-vertex1})-(\ref{Face-vertex4})
\bea &&\sum_{\a=1}^nW^{\a 1}_{1\,\a}(2u)T(\l-N\e_{1}-\e_{i},
\l-\e_{\a}|u)^j_{\a}S(\l-N\e_{1}, \l|u)^{\a}_i\no\\
&&\qquad=\sum_{\a,\b=1}^n
W^{j\,\b}_{\a\,i}(2u)S(\l-N\e_{1}+\e_{\a},
\l+\e_{1}|u)^1_{\b}T(\l-N\e_{1}, \l|u)^{\a}_1.\no\\
\eea Acting both sides on the pseudo-vacuum state
$|vac\rangle^{\l-N\e_{1}}_{\l}$, and using the equations
(\ref{Action-1})-(\ref{Action-2}), we obtain \bea
&&T(\l-N\e_{1}-\e_{i},\l-\e_{1}|u)^j_1S(\l-N\e_{1},\l|u)^1_i
\,|vac\rangle^{\l-N\e_{1}}_{\l}\no\\
&&\qquad=\d^j_i\lt\{
W^{j1}_{\,1j}(2u)-W^{j1}_{\,1j}(2u)\lt(\prod_{k=1}^NW^{1j}_{\,1j}(u-z_k)
W^{j1}_{\,j1}(u+z_k)\rt)\rt\}\no\\
&&\qquad\qquad\qquad\qquad\qquad\times|vac\rangle^{\l-N\e_{1}}_{\l}.\label{C-2}\eea
The equation (\ref{TF-V3}) is a simple consequence of the
equations (\ref{C-1}) and (\ref{C-2}).


\end{document}